\documentclass{article}

\usepackage{microtype}
\usepackage{graphicx}
\usepackage{subfigure}
\usepackage{booktabs} 
\usepackage[inline,shortlabels]{enumitem}
\setlist{itemjoin* = { and\enspace}}
\usepackage{color}
\usepackage{multirow}
\usepackage[figuresright]{rotating}
\usepackage{hyperref}

\usepackage[accepted]{icml2023}

\usepackage{amsmath}
\usepackage{amssymb}
\usepackage{mathtools}
\usepackage{amsthm}

\usepackage[capitalize,noabbrev]{cleveref}

\theoremstyle{plain}

\theoremstyle{definition}

\theoremstyle{remark}

\usepackage[textsize=tiny]{todonotes}

\icmltitlerunning{3D ScatterNet: Inference from 21~cm Light-cones}

\begin{document}

\twocolumn[
\icmltitle{3D ScatterNet: Inference from 21~cm Light-cones}



\icmlsetsymbol{equal}{*}

\begin{icmlauthorlist}
\icmlauthor{Xiaosheng Zhao}{to,too}
\icmlauthor{Shifan Zuo}{tooo,toooo}
\icmlauthor{Yi Mao}{to}
\end{icmlauthorlist}

\icmlaffiliation{to}{Department of Astronomy, Tsinghua University, Beijing, China}
\icmlaffiliation{too}{Institut d'Astrophysique de Paris, CNRS \& Sorbonne Universit\'{e}, Paris, France}
\icmlaffiliation{tooo}{National Astronomical Observatories, Chinese Academy of Sciences, Beijing 100101, China}
\icmlaffiliation{toooo}{Key Laboratory of Radio Astronomy and Technology, Chinese Academy of Sciences, A20 Datun Road, Chaoyang District, Beijing, 100101, P. R. China}


\icmlcorrespondingauthor{Xiaosheng Zhao}{xszhao20@gmail.com}

\icmlkeywords{Machine Learning, ICML}

\vskip 0.3in
]



\printAffiliationsAndNotice{}  

\begin{abstract}
The Square Kilometre Array (SKA) will have the sensitivity to take the 3D light-cones of the 21~cm signal from the epoch of reionization. This signal, however, is highly non-Gaussian and can not be fully interpreted by the traditional statistic using power spectrum. In this work, we introduce the {\tt 3D ScatterNet} that combines the normalizing flows with solid harmonic wavelet scattering transform, a 3D CNN featurizer with inductive bias, to perform implicit likelihood inference (ILI) from 21~cm light-cones. We show that {\tt 3D ScatterNet} outperforms the ILI with a fine-tuned 3D CNN in the literature. It also reaches better performance than ILI with the power spectrum for varied light-cone effects and varied signal contaminations.
\end{abstract}

\section{Introduction}
\label{sec: intro}

The 21~cm fields from the epoch of reionization are highly non-Gaussian resulting from the patch reionization. Maximally exploiting the full information in the 21~cm fields needs new statistics besides the 21~cm power spectrum. Machine learning methods like the convolutional neural networks (CNNs) are promising tools for astrophysical parameter estimation directly from 2D fields or 3D light-cones \citep{Gillet2019,2022MNRAS.509.3852P,2022MNRAS.511.3446N,2021arXiv210503344Z,2022MNRAS.509.3852P}, but some key problems arise: the fine-tuning of hyperparameters and the training process is time-consuming, needs a lot of training data to optimize the learnable parameters, and may lead to sub-optimal trained models \citep{2021arXiv210503344Z,2022MNRAS.509.3852P}.

To solve these problems, people are injecting inductive bias into CNNs with scattering transform \citep{mallat2012group,allys2019rwst,cheng2020new,2022mla..confE..40P} and utilize the scattering transform to construct scattering or wavelet networks \citep{2021arXiv210709539G, 2022mla..confE..40P}. Compared with CNNs, the scattering transform uses filters having well-behaving mathematical structures like the Morlet filters \citep{mallat2012group, 10.1093/mnras/stw1310}. It also uses the special design of modulus nonlinearities, and hierarchical structures, like multi-layers in a CNN, in order to extract the information across scales. With the fixed filter parameters, the scattering transform has the appeal of no need for training, which will be helpful when the training data is scarce and expensive to generate. In 3D applications, the harmonic-related wavelets \citep{eickenberg2017solid,eickenberg2018solid, 2021ApJ...910..122S, 2022PhRvD.105j3534V, 2022MNRAS.517.1625C, 2022arXiv220407646E} are introduced to predict molecular properties or cosmological parameters. Specifically, in \citet{2022arXiv220407646E}, the authors use the first-order wavelet-based features and emphasize the superiority of harmonic wavelets compared with the isotropic and oriented ones. In this work, to extract information from the light cuboids (referred to as light-cones, following the trend in the literature), we utilize the solid harmonic wavelet scattering transform (Solid harmonic WST, \citealp{eickenberg2017solid,eickenberg2018solid}) which injects the inductive bias into 3D CNNs with not only 3D solid harmonic wavelets but the scattering transform which outputs multiple-order wavelet-based features. 

\begin{figure*}[ht]
\vskip 0.2in
\begin{center}
    \centerline{\includegraphics[width=\textwidth]{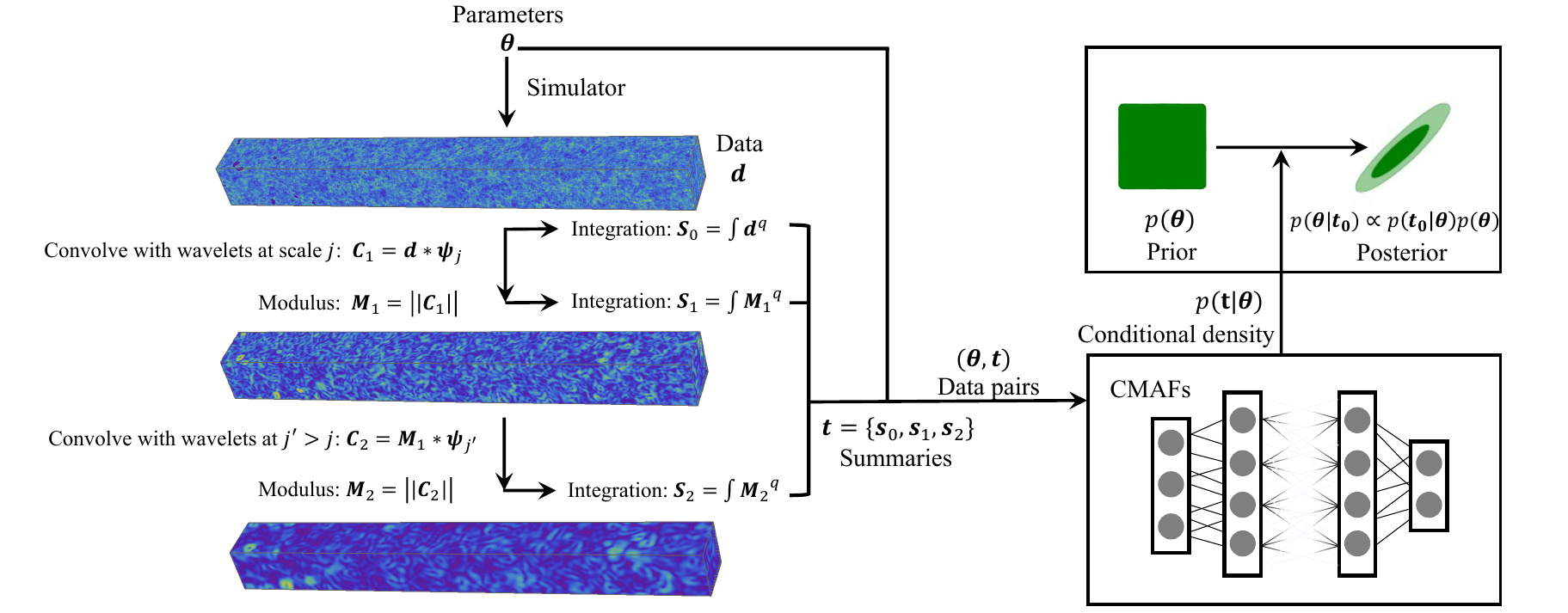}}
    \caption{{\tt 3D ScatterNet}. Left: Solid harmonic WST, the 3D light-cone $\mathbf{d}$ simulated from parameter $\boldsymbol{\theta}$ are compressed by a cascade of scattering transforms, each containing convolution with wavelets, harmonic modulus \citep{eickenberg2017solid}, and integration operation. The integration with raised power $q$ gives the zeroth-, first-, and second-order coefficients $\{\mathbf{S_0}, \mathbf{S_1}, \mathbf{S_2}\}$ which form the summaries $\mathbf{t}$. Right: CMAFs, used to learn the summary density conditional on the parameters, with which the posterior can be inferred at $\mathbf{t_0}$.}
    \label{fig: DELFI-ST}
\end{center}
\vskip -0.2in
\end{figure*}

Conventional CNNs also give only the point estimate of the true parameters, not parameter posteriors, in the Bayesian view. In the case where the likelihood is intractable, the so-called implicit likelihood inference (ILI; \citealp{alsing2018massive,alsing2019fast, 2019arXiv191013233P, Cranmer30055,tejero-cantero2020sbi}) (also called simulation-based inference or likelihood-free inference) is proposed to learn the density of the likelihood or posterior directly from data, with the key methodology like conditional masked autoregressive flows (CMAFs, \citealp{papamakarios2017masked}) which is a kind of normalizing flows \citep{JMLR:v22:19-1028}. In light of this, we construct a {\tt 3D ScatterNet} that combines CMAFs with Solid harmonic WST to infer the parameter posteriors from 21~cm light-cones, compare {\tt 3D ScatterNet} with similar methods using 3D CNN in the literature, and with the conventional statistics 21~cm power spectrum for varied light-cone effects and signal contamination.

\section{Method}
\label{sec: method}

The wavelet used in the Solid harmonic WST is the solid harmonic multiplying by a Gaussian and can be dilated to capture features of the field at different scales. A power $q$ is applied to the modulus operators and results in coefficients that are sensitive to different amplitudes of fields. The covariance properties (both in translation and rotation) can be obtained by aggregating the sub-angle information represented by the angular frequency $m$ within each $l$, and the invariant coefficients can be obtained by integrating over all pixels in the covariant maps. Successive group of wavelets with scales larger than the previous one can be convolved with the previous modulus, in order to get final coefficients that capture the information across scales. Apart from the aforementioned advantage of Solid harmonic WST over 3D CNNs, the ability to apply large kernels easily is also useful to recognize large {\tt HII} bubbles in the 21~cm light-cones; Along the line-of-sight, the wavelets can both retain the local information and preserve some large scale information underneath the long tail of the wavelets. Part of the invariance properties \citep{eickenberg2017solid} of coefficients inherited from the harmonics-the translation invariance over sky directions and the rotation invariance over the redshift axis can benefit the following conditional density learning. 

We calculate the coefficients up to an order of two including the 0th-order coefficients which are defined as the sum of all pixel values raised by the power $q$. We choose the maximum angular frequency number $L=6$ so that $l\in\{0..L\}$ leads to both the solid harmonic wavelets $(l>0)$ which sample angular frequencies capable of decoding underlying structures like filaments, and Gaussian wavelets $(l=0)$ that may characterize the ionizing process. We also choose the maximum scale $J=5$, the modulus power $q\in\{0.5,1,2\}$ for the first and second-order coefficients, and the (half) width parameter ``sigma'' being 1. For these two orders of coefficients, we average the information over different $l$ for each $q$. For the zeroth-order coefficients, $q=0.5$ can lead to the complex number when the integration is negative and $q=1$ is simply zero, so we choose three higher modulus power $q\in\{2,3,4\}$. The final data summary is concatenated by these coefficients flattened over the $q$ and $J$ axes (except the zeroth-order coefficients which have no $j$ dependence) and has a dimension of $66$ which can be decreased by need. We follow \citet{allys2019rwst} and calculate the logarithms (base 2) of these coefficients before averaging (for negative components, we perform the logarithms on their absolute values while keeping their signs). In this work, the Solid harmonic WST is implemented with {\tt Kymatio} \citep{2018arXiv181211214A} and calculated on $66\times66\times660$ or $66^3$ grids depending on the light-cone dimensions.

The features extracted by the Solid harmonic WST serve as the input of CMAFs which aims to perform the implicit likelihood inference, as shown in Fig.~\ref{fig: DELFI-ST}. For CMAFs, we set two neural layers of a single transform, 50 neurons per layer. We also use the ensembles of CMAFs to improve the performance, The number of transforms and the details of ensembles are chosen based on the performance of posterior validation \citep{2021arXiv210503344Z} where the hypothesis tests are adopted to check, statistically, if the posteriors from CMAFs are self-consistent. Note that some works refer to this kind of validation as the calibration, though we do not use this method to calibrate the trained networks. Instead, we use it to indicate if the network complexity and training data are enough to learn the conditional density accurately in a statistical way. The CMAFs are implemented with {\tt pydelfi} \citep{2018arXiv180507226P,lueckmann2018likelihood, alsing2018massive, alsing2019fast} and the CMAFs architectures and training details used in this paper can be found in Appendix~\ref{sec: details}.

\begin{figure*}[ht]
\vskip 0.2in
\begin{center}
    \centerline{\includegraphics[width=\textwidth]{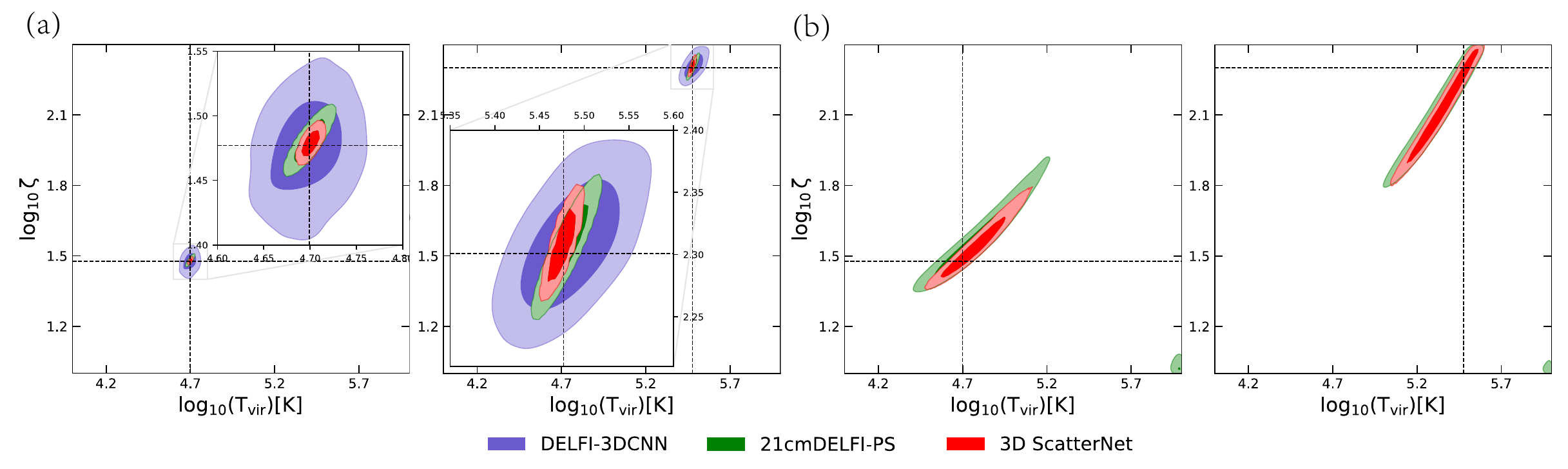}}
    \caption{The posteriors inferred from the (a) pure signal and (b) noised signal, each with the same two representative models. We show the median (cross), $1\sigma$ (dark), and $2\sigma$ (light) confidence regions. The dashed lines indicate the true parameter values. For the pure signal, {\tt 3D ScatterNet} has significant improvement over {\tt DELFI-3DCNN} and has $1.2 - 2$ times improvement over {\tt 21cmDELFI-PS} on $1 \sigma$ marginal errors; For the noised signal, the gains of {\tt 3D ScatterNet} is decreased but still non-negligible ($1.2$ times on average). } 
    \label{fig: comp_pure}
\end{center}
\vskip -0.2in
\end{figure*}

\section{Data}
\label{sec: data}
In this work, we use the publicly available code {\tt 21cmFAST} \citep{Mesinger2007,Mesinger2011}, which can be used to perform semi-numerical simulations of reionization, as the simulator to generate our 21~cm brightness temperature datasets. Following the interpolation approach in \citet{2021arXiv210503344Z}, we generate a light-cone of the size $100\times 100 \times 1000$ comoving ${\rm Mpc}^3$ (or $66\times 66 \times 660$ grid cells) in the redshift range $7.51 \le z \le 11.76$. We parametrize our reionization model as follows: $T_ { \mathrm { vir } }$, the minimum virial temperature of halos that host ionizing sources. We vary this parameter as $ 4 \le \log _ { 10 } \left( T_{ \mathrm { vir } } / \mathrm { K } \right) \le 6 $; $\zeta$, the ionizing efficiency, varied as $1.0 \le \log_{10}(\zeta) \le 2.4 $. We use the logarithmic parameters for training, validation, and testing unless stated otherwise. For the signal with the first level of contamination, dubbed ``pure signal'', we simply remove the mode with ${\bf k}_\perp = 0$, because radio interferometers cannot measure this mode. For the signal with the second level of contamination, dubbed ``noised signal'', we consider 1000-hour SKA1-Low observation of the 3D 21~cm light-cones and use {\tt Tools21cm} \citep{Giri2020} to produce the instrumental thermal noise, where we assume 6-hour observation per day and 10 seconds of integration time. The SKA {\tt uv} coverage is calculated at each frequency channel (out of 660) and used to generate the telescope response on images and suppress the thermal noise. We also smooth the images using a 1-km baseline to increase the signal-to-noise ratio. We also use {\tt pygsm} to simulate the foregrounds based on the GSM-building model \citep{zheng2017improved}, and use the singular value decomposition (SVD, \citealp{masui2013measurement}) for foreground removal, where we regard the first 6 components in the singular matrix as the foreground.

\section{Results}
\label{sec: results}

\textbf{Pure signal}: For the pure signal, we use 18,000 samples for the training and validation of CMAFs. The results for two representative models are shown in Fig.~\ref{fig: comp_pure} panel (a). We find that the results from {\tt 3D ScatterNet} have a significant improvement over that from {\tt DELFI-3DCNN} which is quoted from \citet{2021arXiv210503344Z}. In that work, the authors applied ILI but with data summaries with equal dimensions to the parameters compressed from a trained 3D CNN. 

Note that in the reference paper, the authors used the same simulation settings as ours, but with 9000 samples for training and validation of the 3D CNN and 9000 samples for training and validation of the density estimators. As we repeat with the same experimental settings and increase the training sample size to 18,000 for both 3D CNN and density estimators, we did not get a better performance for these two fiducial models, implying that the performance is mainly limited by the hyper-parameter (including the network architecture) choice in 3D CNN. We also construct a 3D residual network (ResNet) to train a data compressor. However, our limited number of tests show worse performance than the 3D CNN used here. Further fine-tuning efforts are left to future works.

We also show the comparison with {\tt 21cmDELFI-PS} which is quoted from \citet{zhao21cmdelfi}. In {\tt 21cmDELFI-PS}, the authors perform ILI with the power spectrum as the summary that is a concatenated vector (with a dimension of 130) from 10 cubic light-cone boxes, where for each box, the authors choose to
group the modes in Fourier space into 13 k-bins. Comparing {\tt 3D ScatterNet} with {\tt 21cmDELFI-PS}, the estimated $1 \sigma$ marginal errors in the former are $1.2 - 2\times$ smaller than in the latter. In the following, we focus on the comparison of these two methods.

\begin{table}[t]
\caption{Statistical quantities from 300 testing samples: the coefficient of determination $\mathrm { R } ^ { \mathrm { 2 } }$ (calculated in logarithmic parameter space) and the $68\%$ confidence interval of the fractional errors $\epsilon = (  y _ { \mathrm { pred } } - y _ { \mathrm { true } }) / y _ { \mathrm { true } }$ (with unit $\times10^{-1}$, calculated in original parameter space). Both are based on medians of the inferred posteriors and show $ T_{\rm vir}$ and $\zeta$ ordered in rows. \{{\tt PS}, {\tt ST}\} represent \{{\tt 21cmDELFI-PS, {\tt 3D ScatterNet}}\}. {\tt 3D ScatterNet} shows superiority on both quantities for both pure and noised signals.}
\vskip 0.1in
\begin{center}
\begin{small}
\begin{sc}
		\label{tab: perf_comparison}

		\begin{tabular}{lcccc}
			\hline \hline
			&\multicolumn{2}{c}{Pure signal} &\multicolumn{2}{c}{Noised signal} \\
					 \cmidrule(l{.75em}r{.75em}){2-3}
					 \cmidrule(l{.75em}r{.75em}){4-5}
		  {}  & {\tt {\tt PS}} & {\tt {\tt ST}} & {\tt {\tt PS}} & {\tt {\tt ST}}\\
				\hline
		\multirow{2}{*}{$\mathrm{R^2}$} &  0.9989& $\boldsymbol{0.9997}$ & 0.7981& $\boldsymbol{0.8348}$ \\ 
		{} & 0.9978& $\boldsymbol{0.9990}$ & 0.8028& $\boldsymbol{0.8254}$\\
			\hline
		 \multirow{2}{*}{$\epsilon$} &  $[-0.3, 0.3]$ & $[-0.2, 0.2]$ & $[-2.8, 4.7]$& $[-2.8, 4.1]$  \\ 
		{} & $[-0.4, 0.3]$& $[-0.2, 0.2]$ & $[-2.6, 3.6]$& $[-2.2, 3.2]$ \\
		\hline 
		\end{tabular}

\end{sc}
\end{small}
\end{center}
\vskip -0.1in
\end{table}

\begin{figure}[ht]
\vskip 0.1in
\begin{center}
    \centerline{\includegraphics[width=0.5\textwidth]{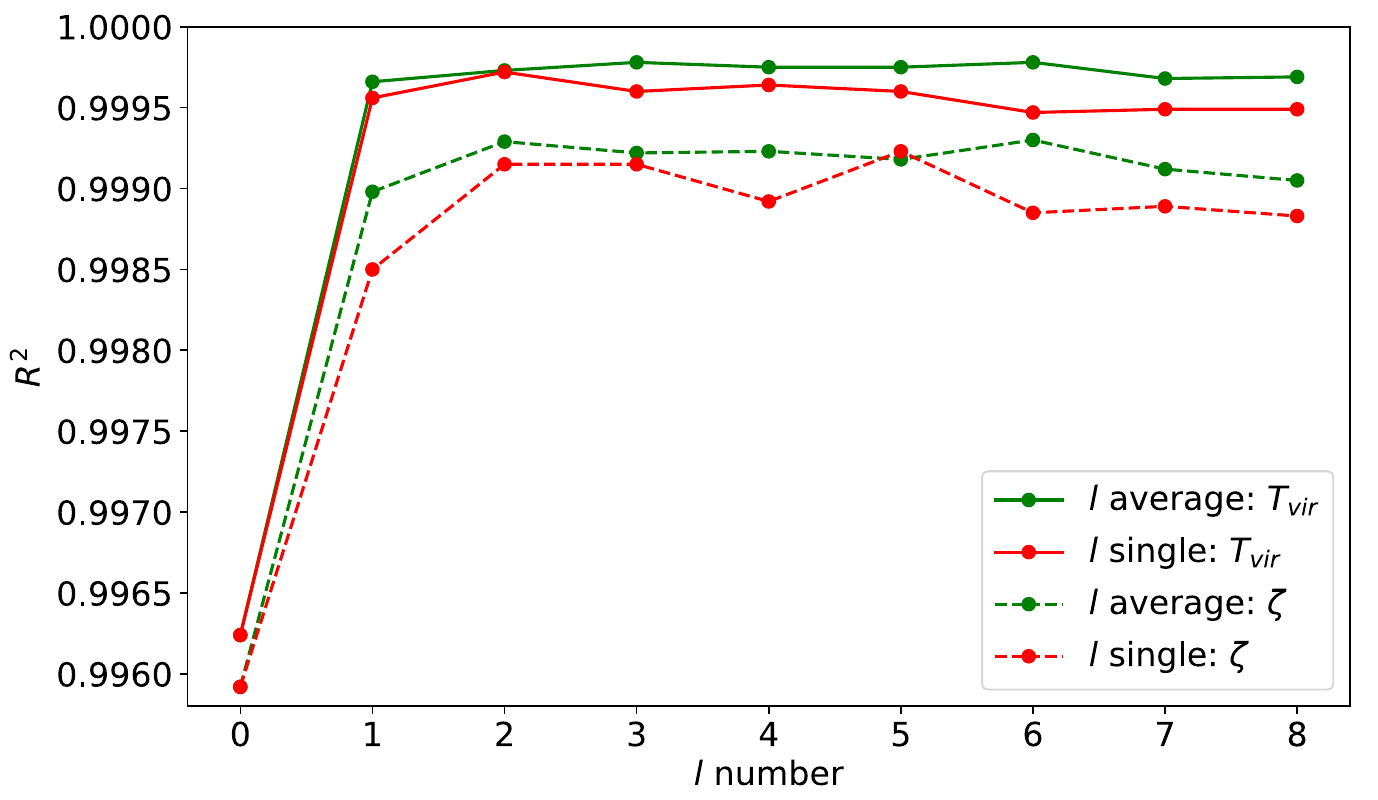}}
    \caption{The $\mathrm{R^2}$ as a function of the angular frequency $l$. The ``$l$ single'' uses the information of a specific $l$ number, while the  ``$l$ average'' uses the average information from less than or equal to a specific $l$ number. It is shown that solid harmonic wavelets ($l>0$) have better performance than Gaussian wavelets ($l=0$). The combination of the two has the best $R^2$.}
    \label{fig:results_l}
\end{center}
\vskip -0.2in
\end{figure}

Next, we test the trained NDEs on 300 samples. In the second row of Table~\ref{tab: perf_comparison}, we present the coefficient of determination $\mathrm{R^2}$ of the logarithmic parameters. A score of $\mathrm { R } ^ { \mathrm { 2 } }$ closer to unity indicates a better overall inference performance of this parameter. We can see both methods give a high $\mathrm { R } ^ { 2 }$ value while {\tt 3D ScatterNet} still outperforms {\tt 21cmDELFI-PS}. In the third row of Table~\ref{tab: perf_comparison}, we present the $68\%$ confidence interval of the fractional errors, $\epsilon = (  y _ { \mathrm { pred } } - y _ { \mathrm { true } }) / y _ { \mathrm { true } }$, where $y$ represents the deduced parameter (in the linear scale), $ T_{\rm vir}$ and $\zeta$, and $y _ { \mathrm { pred } }$ are medians of the inferred posteriors. The fractional errors from{\tt 3D ScatterNet} have about $1.5$ ($1.75$) times improvement in terms of the width of the $68\%$ confidence interval) for $T_{\rm vir}$ ($\zeta$). In Fig.~\ref{fig:results_l}, we show the $\mathrm{R^2}$ of the logarithmic parameters from another two sets of experiments. We first use the information from a single $l$ (``$l$ single''). The $l=0$, corresponding to a Gaussian wavelet, leads to the smallest $R^2$. Then we use the averaged information from different $l$ (``$l$ average''). Compared $l=0$ with other points, the solid harmonic wavelet can lead to better performance than the Gaussian wavelet, and combining the two (by average) has the best $\mathrm{R}^2$. In Table~\ref{tab:single}, we also show the results from light-cubes ($66^3$ grids), where the light-cone effect is less obvious. We adjust the training sample size as shown in Appendix~\ref{sec: details}, in order to get posteriors meeting our validation (calibration) standard. For each specific redshift, {\tt 3D ScatterNet} also has better performance than {\tt 21cmDELFI-PS}. A natural extension of our work is to modify the {\tt 3D ScatterNet} by concatenating the scattering coefficients from discrete boxes, with the caveat that there are much more coefficients and thus it is harder to train the CMAFs.

\begin{table}[t]
\caption{Similar to Table~\ref{tab: perf_comparison} but only shows the $68\%$ confidence interval of the fractional errors $\epsilon$ from light-cubes ($66^3$ grids) for the pure signal. The two light-cubes correspond to the central redshifts \{8.36, 9.11\}. The ``\begin{sc}Full\end{sc}'' represents the full-band light-cones used for the main results of this paper. {\tt 3D ScatterNet(ST)} also shows better performance than {\tt 21cmDELFI-PS(PS)} for light-cubes with less light-cone effects.}
\vskip 0.1in
\begin{center}
\begin{small}
\begin{sc}
    \begin{tabular}{lccc}
    \hline \hline

			  &  Light-cube$_1$ & Light-cube$_2$ & full\\
        \hline
\multirow{2}{*}{$\epsilon_{\mathrm{PS}}$} & [-0.6, 0.8]& [-0.7, 0.8] & [-0.3, 0.3]\\ 
{}  & [-0.5, 0.8]& [-0.8, 0.7]& [-0.4, 0.3]\\
\hline
\multirow{2}{*}{$\epsilon_{\mathrm{ST}}$}  & [-0.3, 0.4] &[-0.5, 0.4]&[-0.1, 0.1]\\ 
{} &[-0.4, 0.5] & [-0.6, 0.4]&[-0.2, 0.2]\\
\hline 
\end{tabular}
\end{sc}
\end{small}
\end{center}
\vskip -0.2in
\label{tab:single}
\end{table}

The results above are shown to be statistically reliable with the validation methods we use. From the hypothesis tests for validation, we claim that our results are at least reliable with a significance of 0.01. We emphasize the importance of using the validation methods to ensure the CMAFs can learn self-consistent posteriors.  The improvement of {\tt 3D ScatterNet} over {\tt 21cmDELFI-PS} implies that the Solid harmonic WST may better capture the inherent non-Gaussianity and the evolution information of the 21~cm light-cones. We visualize the coefficients from Solid harmonic WST in Appendix~\ref{sec: visual} for a better understanding of how these coefficients constrain the astrophysical parameters.

\textbf{Noised signal}: For the noised signal, we use 36,000 samples for the training and validation of CMAFs. Since the images after the smoothing with a 1-km baseline lose the small-scale information, we discard the components of scattering coefficients with $j=0$ and the final coefficients have the dimension of 48. For the power spectrum, its dimension is reasonably reduced to 70 with little information loss. The inference results are shown in Fig.~\ref{fig: comp_pure} panel (b), where we concatenate the MCMC chains from 10 mock observations each with a different realization of thermal noise. The results from {\tt 3D ScatterNet} also have slight improvement compared with that from {\tt 21cmDELFI-PS} in terms of $1\sigma$ marginal error. From the last two columns of Table~\ref{tab: perf_comparison}, the $R^2$ from {\tt 3D ScatterNet} are about $5\%$ ($3\%$) higher for $\log _ { 10 } \left( T_ { \rm vir } \right)$ ($\mathrm{log_{10}\zeta}$) and the fractional errors have about $1.09$ ($1.15$) times improvement for $T_{\rm vir}$ ($\zeta$).

\section{Summary}
\label{sec: summary}

In this work, we build a {\tt 3D ScatterNet} by combining the Solid harmonic WST, a 3D CNN featurizer with inductive bias, with CMAFs in order to perform implicit likelihood inference (ILI) from 21~cm light-cones. After using the posterior validation tools to choose the CMAFs architecture and the proper size of training sets, we find that solid harmonic wavelets can enhance the performance compared with using Gaussian wavelets alone. {\tt 3D ScatterNet} has significant improvement over using 3D CNN for ILI in the literature. Our results show that the Solid harmonic WST produces informative summaries more robustly and efficiently compared with a 3D CNN resulting from a reasonable amount of fine-tuning, while we note that the 3D CNN architecture adopted in this reference paper may not be optimal and could be improved by introducing further inductive bias in the networks or adopting a learnable version of 3D scattering transform. We also find improvements over using 21~cm power spectrum for ILI regarding varied light-cone effects and signal contamination, implying that applying Solid harmonic WST to full-band light-cones better captures the inherent non-Gaussianity and evolution information in line-of-sight, at either small or large scales, without splitting light-cones. Our results show that {\tt 3D ScatterNet} possesses the potential for parameter inference with future 21~cm light-cones and other line intensity mappings.

\section*{Acknowledgements}

This work was supported by the National SKA Program of China (grant No.~2020SKA0110401), NSFC (grant No.~11821303), and National Key R\&D Program of China (grant No.~2018YFA0404502). We thank Yuan-Sen Ting for his kind help in preparing the initial version of the manuscript for submission. We also thank the reviewers for giving useful comments to improve this manuscript. We acknowledge the Tsinghua Astrophysics High-Performance Computing platform at Tsinghua University for providing computational and data storage resources that have contributed to the research results reported within this paper.


\bibliography{main}

\begin{thebibliography}{32}
\providecommand{\natexlab}[1]{#1}
\providecommand{\url}[1]{\texttt{#1}}
\expandafter\ifx\csname urlstyle\endcsname\relax
  \providecommand{\doi}[1]{doi: #1}\else
  \providecommand{\doi}{doi: \begingroup \urlstyle{rm}\Url}\fi

\bibitem[{Allys} et~al.(2019){Allys}, {Levrier}, {Zhang}, {Colling},
  {Regaldo-Saint Blancard}, {Boulanger}, {Hennebelle}, and
  {Mallat}]{allys2019rwst}
{Allys}, E., {Levrier}, F., {Zhang}, S., {Colling}, C., {Regaldo-Saint
  Blancard}, B., {Boulanger}, F., {Hennebelle}, P., and {Mallat}, S.
\newblock {The RWST, a comprehensive statistical description of the
  non-Gaussian structures in the ISM}.
\newblock \emph{\aap}, 629:\penalty0 A115, September 2019.
\newblock \doi{10.1051/0004-6361/201834975}.

\bibitem[{Alsing} et~al.(2018){Alsing}, {Wandelt}, and
  {Feeney}]{alsing2018massive}
{Alsing}, J., {Wandelt}, B., and {Feeney}, S.
\newblock {Massive optimal data compression and density estimation for
  scalable, likelihood-free inference in cosmology}.
\newblock \emph{\mnras}, 477\penalty0 (3):\penalty0 2874--2885, July 2018.
\newblock \doi{10.1093/mnras/sty819}.

\bibitem[{Alsing} et~al.(2019){Alsing}, {Charnock}, {Feeney}, and
  {Wandelt}]{alsing2019fast}
{Alsing}, J., {Charnock}, T., {Feeney}, S., and {Wandelt}, B.
\newblock {Fast likelihood-free cosmology with neural density estimators and
  active learning}.
\newblock \emph{\mnras}, 488\penalty0 (3):\penalty0 4440--4458, Sep 2019.
\newblock \doi{10.1093/mnras/stz1960}.

\bibitem[{Andreux} et~al.(2018){Andreux}, {Angles}, {Exarchakis},
  {Leonarduzzi}, {Rochette}, {Thiry}, {Zarka}, {Mallat}, {And{\'e}n},
  {Belilovsky}, {Bruna}, {Lostanlen}, {Hirn}, {Oyallon}, {Zhang}, {Cella}, and
  {Eickenberg}]{2018arXiv181211214A}
{Andreux}, M., {Angles}, T., {Exarchakis}, G., {Leonarduzzi}, R., {Rochette},
  G., {Thiry}, L., {Zarka}, J., {Mallat}, S., {And{\'e}n}, J., {Belilovsky},
  E., {Bruna}, J., {Lostanlen}, V., {Hirn}, M.~J., {Oyallon}, E., {Zhang}, S.,
  {Cella}, C., and {Eickenberg}, M.
\newblock {Kymatio: Scattering Transforms in Python}.
\newblock \emph{arXiv e-prints}, art. arXiv:1812.11214, December 2018.

\bibitem[{Cheng} et~al.(2020){Cheng}, {Ting}, {M{\'e}nard}, and
  {Bruna}]{cheng2020new}
{Cheng}, S., {Ting}, Y.-S., {M{\'e}nard}, B., and {Bruna}, J.
\newblock {A new approach to observational cosmology using the scattering
  transform}.
\newblock \emph{\mnras}, 499\penalty0 (4):\penalty0 5902--5914, December 2020.
\newblock \doi{10.1093/mnras/staa3165}.

\bibitem[{Chung}(2022)]{2022MNRAS.517.1625C}
{Chung}, D.~T.
\newblock {Exploration of 3D wavelet scattering transform coefficients for
  line-intensity mapping measurements}.
\newblock \emph{\mnras}, 517\penalty0 (2):\penalty0 1625--1639, December 2022.
\newblock \doi{10.1093/mnras/stac2662}.

\bibitem[Cranmer et~al.(2020)Cranmer, Brehmer, and Louppe]{Cranmer30055}
Cranmer, K., Brehmer, J., and Louppe, G.
\newblock The frontier of simulation-based inference.
\newblock \emph{Proceedings of the National Academy of Sciences}, 117\penalty0
  (48):\penalty0 30055--30062, 2020.
\newblock ISSN 0027-8424.
\newblock \doi{10.1073/pnas.1912789117}.
\newblock URL \url{https://www.pnas.org/content/117/48/30055}.

\bibitem[Eickenberg et~al.(2017)Eickenberg, Exarchakis, Hirn, and
  Mallat]{eickenberg2017solid}
Eickenberg, M., Exarchakis, G., Hirn, M., and Mallat, S.
\newblock Solid harmonic wavelet scattering: Predicting quantum molecular
  energy from invariant descriptors of 3d electronic densities.
\newblock In \emph{Proceedings of the 31st International Conference on Neural
  Information Processing Systems}, NIPS'17, pp.\  6543–6552, Red Hook, NY,
  USA, 2017. Curran Associates Inc.
\newblock ISBN 9781510860964.

\bibitem[Eickenberg et~al.(2018)Eickenberg, Exarchakis, Hirn, Mallat, and
  Thiry]{eickenberg2018solid}
Eickenberg, M., Exarchakis, G., Hirn, M., Mallat, S., and Thiry, L.
\newblock Solid harmonic wavelet scattering for predictions of molecule
  properties.
\newblock \emph{The Journal of chemical physics}, 148\penalty0 (24):\penalty0
  241732, 2018.

\bibitem[{Eickenberg} et~al.(2022){Eickenberg}, {Allys}, {Moradinezhad Dizgah},
  {Lemos}, {Massara}, {Abidi}, {Hahn}, {Hassan}, {Regaldo-Saint Blancard},
  {Ho}, {Mallat}, {And{\'e}n}, and {Villaescusa-Navarro}]{2022arXiv220407646E}
{Eickenberg}, M., {Allys}, E., {Moradinezhad Dizgah}, A., {Lemos}, P.,
  {Massara}, E., {Abidi}, M., {Hahn}, C., {Hassan}, S., {Regaldo-Saint
  Blancard}, B., {Ho}, S., {Mallat}, S., {And{\'e}n}, J., and
  {Villaescusa-Navarro}, F.
\newblock {Wavelet Moments for Cosmological Parameter Estimation}.
\newblock \emph{arXiv e-prints}, art. arXiv:2204.07646, April 2022.
\newblock \doi{10.48550/arXiv.2204.07646}.

\bibitem[{Gauthier} et~al.(2021){Gauthier}, {Th{\'e}rien},
  {Als{\`e}ne-Racicot}, {Chaudhary}, {Rish}, {Belilovsky}, {Eickenberg}, and
  {Wolf}]{2021arXiv210709539G}
{Gauthier}, S., {Th{\'e}rien}, B., {Als{\`e}ne-Racicot}, L., {Chaudhary}, M.,
  {Rish}, I., {Belilovsky}, E., {Eickenberg}, M., and {Wolf}, G.
\newblock {Parametric Scattering Networks}.
\newblock \emph{arXiv e-prints}, art. arXiv:2107.09539, July 2021.
\newblock \doi{10.48550/arXiv.2107.09539}.

\bibitem[{Gillet} et~al.(2019){Gillet}, {Mesinger}, {Greig}, {Liu}, and
  {Ucci}]{Gillet2019}
{Gillet}, N., {Mesinger}, A., {Greig}, B., {Liu}, A., and {Ucci}, G.
\newblock {Deep learning from 21-cm tomography of the cosmic dawn and
  reionization}.
\newblock \emph{\mnras}, 484:\penalty0 282--293, March 2019.
\newblock \doi{10.1093/mnras/stz010}.

\bibitem[Giri et~al.(2020)Giri, Mellema, and Jensen]{Giri2020}
Giri, S.~K., Mellema, G., and Jensen, H.
\newblock Tools21cm: A python package to analyse the large-scale 21-cm signal
  from the epoch of reionization and cosmic dawn.
\newblock \emph{Journal of Open Source Software}, 5\penalty0 (52):\penalty0
  2363, 2020.
\newblock \doi{10.21105/joss.02363}.
\newblock URL \url{https://doi.org/10.21105/joss.02363}.

\bibitem[Lueckmann et~al.(2019)Lueckmann, Bassetto, Karaletsos, and
  Macke]{lueckmann2018likelihood}
Lueckmann, J.-M., Bassetto, G., Karaletsos, T., and Macke, J.~H.
\newblock Likelihood-free inference with emulator networks.
\newblock In Ruiz, F., Zhang, C., Liang, D., and Bui, T. (eds.),
  \emph{Proceedings of The 1st Symposium on Advances in Approximate Bayesian
  Inference}, volume~96 of \emph{Proceedings of Machine Learning Research},
  pp.\ ~32, Montreal, 02 Dec 2019. PMLR.

\bibitem[Mallat(2012)]{mallat2012group}
Mallat, S.
\newblock Group invariant scattering.
\newblock \emph{Communications on Pure and Applied Mathematics}, 65\penalty0
  (10):\penalty0 1331--1398, 2012.
\newblock \doi{https://doi.org/10.1002/cpa.21413}.
\newblock URL \url{https://onlinelibrary.wiley.com/doi/abs/10.1002/cpa.21413}.

\bibitem[{Masui} et~al.(2013){Masui}, {Switzer}, {Banavar}, {Bandura}, {Blake},
  {Calin}, {Chang}, {Chen}, {Li}, {Liao}, {Natarajan}, {Pen}, {Peterson},
  {Shaw}, and {Voytek}]{masui2013measurement}
{Masui}, K.~W., {Switzer}, E.~R., {Banavar}, N., {Bandura}, K., {Blake}, C.,
  {Calin}, L.~M., {Chang}, T.~C., {Chen}, X., {Li}, Y.~C., {Liao}, Y.~W.,
  {Natarajan}, A., {Pen}, U.~L., {Peterson}, J.~B., {Shaw}, J.~R., and
  {Voytek}, T.~C.
\newblock {Measurement of 21 cm Brightness Fluctuations at z
  \raisebox{-0.5ex}\textasciitilde 0.8 in Cross-correlation}.
\newblock \emph{\apjl}, 763\penalty0 (1):\penalty0 L20, January 2013.
\newblock \doi{10.1088/2041-8205/763/1/L20}.

\bibitem[{Mesinger} \& {Furlanetto}(2007){Mesinger} and
  {Furlanetto}]{Mesinger2007}
{Mesinger}, A. and {Furlanetto}, S.
\newblock {Efficient Simulations of Early Structure Formation and
  Reionization}.
\newblock \emph{\apj}, 669\penalty0 (2):\penalty0 663--675, November 2007.
\newblock \doi{10.1086/521806}.

\bibitem[Mesinger et~al.(2011)Mesinger, Furlanetto, and Cen]{Mesinger2011}
Mesinger, A., Furlanetto, S., and Cen, R.
\newblock 21cmfast: a fast, seminumerical simulation of the high-redshift 21-cm
  signal.
\newblock \emph{\mnras}, 411\penalty0 (2):\penalty0 955--972, 2011.
\newblock \doi{10.1111/j.1365-2966.2010.17731.x}.
\newblock URL \url{http://dx.doi.org/10.1111/j.1365-2966.2010.17731.x}.

\bibitem[{Neutsch} et~al.(2022){Neutsch}, {Heneka}, and
  {Br{\"u}ggen}]{2022MNRAS.511.3446N}
{Neutsch}, S., {Heneka}, C., and {Br{\"u}ggen}, M.
\newblock {Inferring astrophysics and dark matter properties from 21 cm
  tomography using deep learning}.
\newblock \emph{\mnras}, 511\penalty0 (3):\penalty0 3446--3462, April 2022.
\newblock \doi{10.1093/mnras/stac218}.

\bibitem[{Papamakarios}(2019)]{2019arXiv191013233P}
{Papamakarios}, G.
\newblock {Neural Density Estimation and Likelihood-free Inference}.
\newblock \emph{arXiv e-prints}, art. arXiv:1910.13233, October 2019.

\bibitem[Papamakarios et~al.(2017)Papamakarios, Pavlakou, and
  Murray]{papamakarios2017masked}
Papamakarios, G., Pavlakou, T., and Murray, I.
\newblock Masked autoregressive flow for density estimation.
\newblock In \emph{Proceedings of the 31st International Conference on Neural
  Information Processing Systems}, NIPS'17, pp.\  2335, Red Hook, NY, USA,
  2017. Curran Associates Inc.
\newblock ISBN 9781510860964.

\bibitem[{Papamakarios} et~al.(2018){Papamakarios}, {Sterratt}, and
  {Murray}]{2018arXiv180507226P}
{Papamakarios}, G., {Sterratt}, D.~C., and {Murray}, I.
\newblock {Sequential Neural Likelihood: Fast Likelihood-free Inference with
  Autoregressive Flows}.
\newblock \emph{arXiv e-prints}, art. arXiv:1805.07226, May 2018.
\newblock \doi{10.48550/arXiv.1805.07226}.

\bibitem[Papamakarios et~al.(2021)Papamakarios, Nalisnick, Rezende, Mohamed,
  and Lakshminarayanan]{JMLR:v22:19-1028}
Papamakarios, G., Nalisnick, E., Rezende, D.~J., Mohamed, S., and
  Lakshminarayanan, B.
\newblock Normalizing flows for probabilistic modeling and inference.
\newblock \emph{Journal of Machine Learning Research}, 22\penalty0
  (57):\penalty0 1--64, 2021.
\newblock URL \url{http://jmlr.org/papers/v22/19-1028.html}.

\bibitem[{Pedersen} et~al.(2022){Pedersen}, {Ho}, and
  {Eickenberg}]{2022mla..confE..40P}
{Pedersen}, C., {Ho}, S., and {Eickenberg}, M.
\newblock {Learnable wavelet neural networks for cosmological inference}.
\newblock In \emph{Machine Learning for Astrophysics}, pp.\ ~40, July 2022.

\bibitem[{Prelogovi{\'c}} et~al.(2022){Prelogovi{\'c}}, {Mesinger}, {Murray},
  {Fiameni}, and {Gillet}]{2022MNRAS.509.3852P}
{Prelogovi{\'c}}, D., {Mesinger}, A., {Murray}, S., {Fiameni}, G., and
  {Gillet}, N.
\newblock {Machine learning astrophysics from 21 cm lightcones: impact of
  network architectures and signal contamination}.
\newblock \emph{\mnras}, 509\penalty0 (3):\penalty0 3852--3867, January 2022.
\newblock \doi{10.1093/mnras/stab3215}.

\bibitem[{Saydjari} et~al.(2021){Saydjari}, {Portillo}, {Slepian}, {Kahraman},
  {Burkhart}, and {Finkbeiner}]{2021ApJ...910..122S}
{Saydjari}, A.~K., {Portillo}, S. K.~N., {Slepian}, Z., {Kahraman}, S.,
  {Burkhart}, B., and {Finkbeiner}, D.~P.
\newblock {Classification of Magnetohydrodynamic Simulations Using Wavelet
  Scattering Transforms}.
\newblock \emph{\apj}, 910\penalty0 (2):\penalty0 122, April 2021.
\newblock \doi{10.3847/1538-4357/abe46d}.

\bibitem[Tejero-Cantero et~al.(2020)Tejero-Cantero, Boelts, Deistler,
  Lueckmann, Durkan, Gonçalves, Greenberg, and Macke]{tejero-cantero2020sbi}
Tejero-Cantero, A., Boelts, J., Deistler, M., Lueckmann, J.-M., Durkan, C.,
  Gonçalves, P.~J., Greenberg, D.~S., and Macke, J.~H.
\newblock sbi: A toolkit for simulation-based inference.
\newblock \emph{Journal of Open Source Software}, 5\penalty0 (52):\penalty0
  2505, 2020.
\newblock \doi{10.21105/joss.02505}.
\newblock URL \url{https://doi.org/10.21105/joss.02505}.

\bibitem[Trott(2016)]{10.1093/mnras/stw1310}
Trott, C.~M.
\newblock {Exploring the evolution of reionization using a wavelet transform
  and the light cone effect}.
\newblock \emph{\mnras}, 461\penalty0 (1):\penalty0 126--135, 06 2016.
\newblock ISSN 0035-8711.
\newblock \doi{10.1093/mnras/stw1310}.
\newblock URL \url{https://doi.org/10.1093/mnras/stw1310}.

\bibitem[{Valogiannis} \& {Dvorkin}(2022){Valogiannis} and
  {Dvorkin}]{2022PhRvD.105j3534V}
{Valogiannis}, G. and {Dvorkin}, C.
\newblock {Towards an optimal estimation of cosmological parameters with the
  wavelet scattering transform}.
\newblock \emph{\prd}, 105\penalty0 (10):\penalty0 103534, May 2022.
\newblock \doi{10.1103/PhysRevD.105.103534}.

\bibitem[{Zhao} et~al.(2022{\natexlab{a}}){Zhao}, {Mao}, {Cheng}, and
  {Wandelt}]{2021arXiv210503344Z}
{Zhao}, X., {Mao}, Y., {Cheng}, C., and {Wandelt}, B.~D.
\newblock {Simulation-based Inference of Reionization Parameters from 3D
  Tomographic 21 cm Light-cone Images}.
\newblock \emph{\apj}, 926\penalty0 (2):\penalty0 151, February
  2022{\natexlab{a}}.
\newblock \doi{10.3847/1538-4357/ac457d}.

\bibitem[{Zhao} et~al.(2022{\natexlab{b}}){Zhao}, {Mao}, and
  {Wandelt}]{zhao21cmdelfi}
{Zhao}, X., {Mao}, Y., and {Wandelt}, B.~D.
\newblock {Implicit Likelihood Inference of Reionization Parameters from the 21
  cm Power Spectrum}.
\newblock \emph{arXiv e-prints}, art. arXiv:2203.15734, March
  2022{\natexlab{b}}.

\bibitem[{Zheng} et~al.(2017){Zheng}, {Tegmark}, {Dillon}, {Kim}, {Liu},
  {Neben}, {Jonas}, {Reich}, and {Reich}]{zheng2017improved}
{Zheng}, H., {Tegmark}, M., {Dillon}, J.~S., {Kim}, D.~A., {Liu}, A., {Neben},
  A.~R., {Jonas}, J., {Reich}, P., and {Reich}, W.
\newblock {An improved model of diffuse galactic radio emission from 10 MHz to
  5 THz}.
\newblock \emph{\mnras}, 464\penalty0 (3):\penalty0 3486--3497, January 2017.
\newblock \doi{10.1093/mnras/stw2525}.

\end{thebibliography}
\bibliographystyle{icml2023}


\newpage
\appendix
\onecolumn

\section{CMAFs architectures and training details}
\label{sec: details}
In {\tt pydelfi}, we set two neural layers of a single transform, 50 neurons per layer, for all CMAFs architectures. We also use the ensembles of CMAFs which are shown to be more effective compared with the single CMAF for small size of the training sample. For the training of CMAFs, we split and use $10\%$ of the training samples for additional validation. We also use the batch size of 50, epochs of 2000, and patience of 20 for early stopping. We fine-tune the the size of training sample and the CMAFs architecture for each method and each experiment, after the results of posterior validation (calibration). The general idea for the fine-tuning is that the data (summaries) containing more uncertainties may need more complex CMAFs like more transformations in a single CMAF, and the data (summaries) with a higher dimension may need a larger size of training sample because of the huge feature space. The details to produce the results in this paper are as follows:

{\tt 3D ScatterNet}:
\begin{itemize}
    \item Pure signal (main results of light-cones). We use 18,000 samples and an ensemble of 8 CMAFs: $(5,6,7,8)*2$, where we use two CMAFs blocks each containing a CMAF with the number of transformations 5, 6, 7, and 8, respectively. Hereafter we use a similar convention for the illustration of CMAFs ensembles. We also find that the performance can be further enhanced slightly especially for {\tt 3D ScatterNet} with double size of training samples.
    \item Pure signal (light-cones with the information of averaged $l$). We use the ensembles from $l=1$ to $l=8$: $(5)*4$, $(6,7,8,9)*3$, $(6,7,8,9)*2$, $(5,6,7,8)*2$, $(6,7,8,9)*3$, $(5,6,7,8)*2$, $(5)*4$, $(5,6,7,8)*2$.
    \item Pure signal (light-cones with the information of single $l$). We use the ensembles from $l=0$ to $l=8$: $(5)*4$, $(5)*4$, $(5,6,7,8)*2$, $(5)*4$, $(5,6,7,8)*2$, $(5,6,7,8)*2$, $(5,6,7,8)*2$, $(5,6,7,8)*2$, $(5,6,7,8)*2$.
    \item Pure signal (light-cubes). We use 27,000 samples and the ensembles for the 1st, 3rd, and 5th box: $(5,6,7,8)*3$, $(5,6,7,8)*3$, $(5,6,7,8)*2$.
    \item Noised signal. We use 36,000 samples and the ensembles: $(20)*2$.
\end{itemize}
{\tt 21cmDELFI-PS}:
\begin{itemize}
    \item Pure signal (main results of light-cones). We use 18,000 samples and the ensembles $(5)*4$.
    \item Pure signal (light-cubes). For the 1st box, we use 32,596 samples and the ensembles $(6,7,8,9)*3$; for the 3rd box, we use 34,044 samples and the ensembles $(5,6,7,8)*3$; and for the 5th box we use 27,000 samples and the ensembles $(5,6,7,8)*2$.
    \item Noised signal. We use 36,000 samples and the ensembles $(20)*3$.
\end{itemize}

\section{The sensitively of scattering coefficients to astrophysical parameters}
\label{sec: visual}
\begin{figure*}[ht]
\vskip 0.2in
\begin{center}
    \centerline{\includegraphics[width=\textwidth]{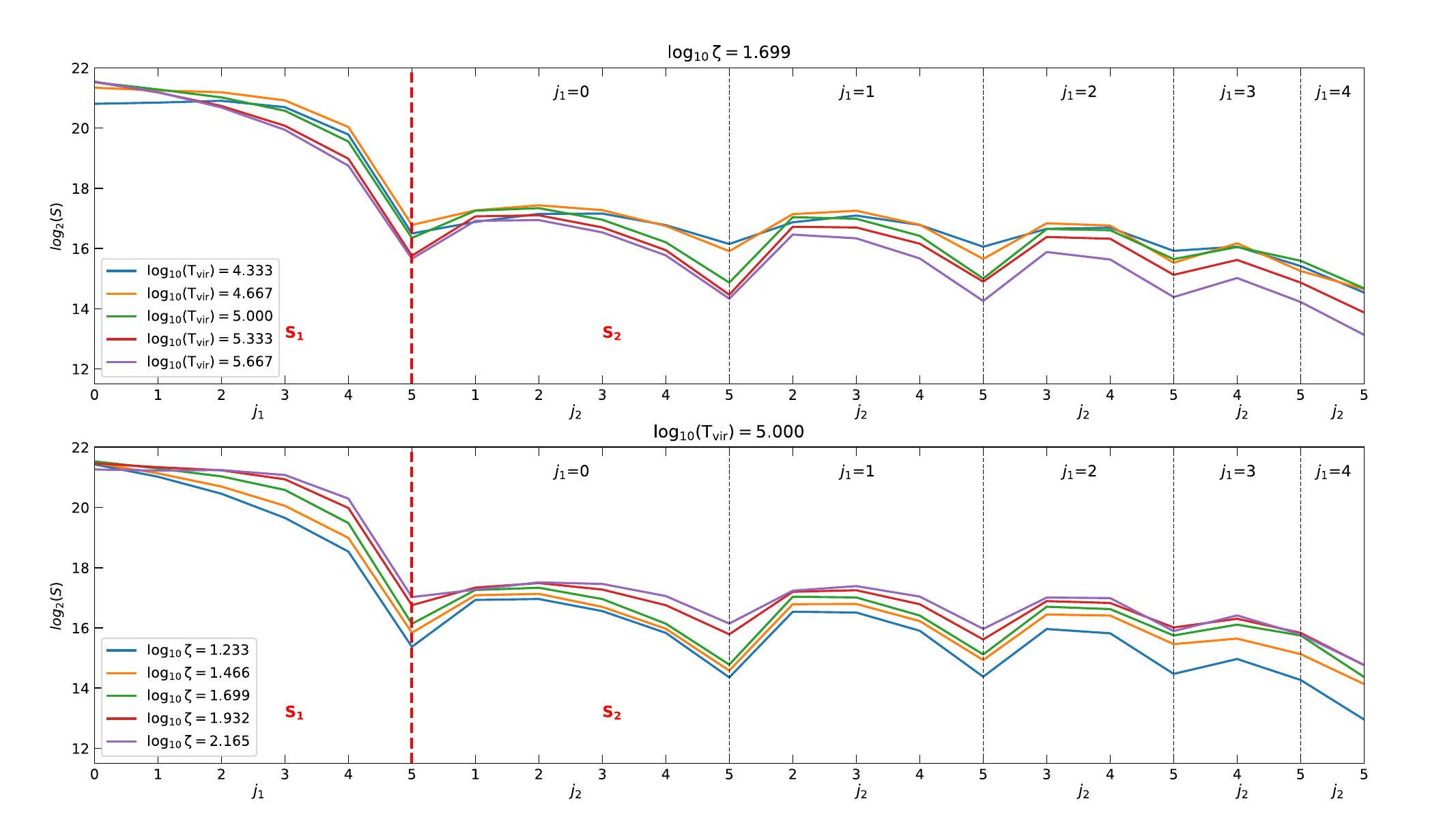}}
    \caption{The first ($\mathbf{S_1}$) and second ($\mathbf{S_2}$) order solid harmonic scattering (logarithmic) coefficients with averaged $l$ (up to $l=6$) and $q=1$. We vary the two parameters $\log _ { 10 } \left( T_ { \rm vir } \right)$ and $\mathrm{log_{10}\zeta}$ in the top and bottom panel, respectively. The green lines in both the top and bottom panels are the same and are used for a clearer comparison.}
    \label{fig:statistics}
\end{center}
\vskip -0.2in
\end{figure*}

We show the first- and second-order scattering coefficients calculated from the light-cones (pure signal) in Fig.~\ref{fig:statistics} with averaged $l$ (up to $l=6$) and $q=1$. We find that increasing $\log _ { 10 } \left( T_ { \rm vir } \right)$ ($\mathrm{log_{10}\zeta}$) tends to have decreasing (increasing) amplitude of the coefficients at all scales. The sensitivity of the amplitude to the varying parameters implies that these coefficients have the power to constrain the two parameters. The opposite effects on the amplitude agree with our physical intuition: perturbing a specific universe with larger ionizing efficiency has a similar ionization history to perturbing with more abundant low-mass galaxies, which also implies the direction of degeneracies in the two-parameter space. Interestingly, our initial results show that only using the zeroth-order coefficients can already give the inference performance just slightly worse than {\tt 21cmDELFI-PS} on one of the representative models, while it failed sampling of the posterior with the default settings of {\tt emcee} for the other model. Further investigation is needed based on these results.

\end{document}